\documentclass[preprint]{elsarticle}

\title{Tag Spotting at the interference range}

\usepackage{savesym}

\usepackage{amsmath}
\usepackage{amssymb}
\usepackage{amsfonts}

\usepackage{graphicx}

\usepackage[caption=false,font=footnotesize]{subfig}
\usepackage{url}
\usepackage{array}
\usepackage{verbatim}

\newlength{\myindent}
\usepackage{color}
\definecolor{red}{rgb}{1,0,0}
\definecolor{blue}{rgb}{0,0,1}

\newcommand{\vlad}[1]{{#1}}

\newcommand{\tocut}[1]{}
\newcommand{\mindent}[0]{{\hspace{\myindent}}}

\begin{document}

\begin{frontmatter}

\author[usc]{Horia Vlad Balan}
\ead{hbalan@usc.edu}

\author[usc]{Konstantinos Psounis\corref{cor1}}
\ead{kpsounis@usc.edu}

\address[usc]{University of Southern California, 3740 McClintock Ave,
  Los Angeles, CA 90089, USA}

\cortext[cor1]{Corresponding author: Phone: +1 213 740 4413}

\begin{abstract}

In wireless networks, the presence of interference among wireless
links introduces dependencies among flows that do not share a single
link or node. As a result, when designing a resource allocation
scheme, be it a medium access scheduler or a flow rate controller, one
needs to consider the interdependence among nodes within interference
range of each other. Specifically, control plane information needs to
reach nearby nodes which often lie outside the communication range,
but within the interference range of a node of interest.
  
But how can one communicate control plane information well beyond the
existing communication range? To address this fundamental need we
introduce tag spotting. Tag spotting refers to a communication system
which allows reliable control data transmission at SNR values as low
as 0 dB. It does this by employing a number of signal encoding
techniques including adding redundancy to multitone modulation,
shaping the spectrum to reduce inter-carrier interference, and the use
of algebraic coding.  Making use of a detection theory-based model we
analyze the performance achievable by our modulation as well as the
trade-off between the rate of the information transmitted and the
likelihood of error.  Using real-world experiments on an OFDM system
built with software radios, we show that we can transmit data at the
target SNR value of 0 dB with a 6\% overhead; that is, 6\% of our
packet is used for our low-SNR decodable tags (which carry up to a
couple of bytes of data in our testbed), while the remaining 94\% is
used for traditional header and payload data. We also demonstrate via
simulations how tag spotting can be used in implementing fair and
efficient rate control and scheduling schemes in the context of
wireless multi-hop networks, while pointing out that the idea of tag
spotting is useful in the context of any wireless network in which
control-plane information must travel beyond the communication range
of a node.

\end{abstract}

\maketitle


\end{frontmatter}

\section{Introduction}
\label{section:introduction}

Many of the challenges encountered in the design of wireless networks
with multiple transmission and reception points stem from the quirks
of wireless signal propagation. Using currently prevailing
transmission techniques, wireless signals cannot be focused
exclusively towards their intended recipient, making wireless an
inherently shared medium. Wireless transmissions are local in their
coverage, and, in general, no sender or receiver will have access to
complete channel state information.  Because of these characteristics,
a wireless network is commonly modeled as a set of links among which
interference may occur depending on the particular choice of senders
transmitting at the same time. The effects of wireless interference
are far reaching, affecting all network layers, from physical layer
and medium access to flow control and user satisfaction. They extend
beyond the space of a single host or a single link, as flows that do
not share any hosts or links in their paths might in fact find
themselves competing for resources.  Its direct consequence is
unfairness leading to flow starvation and underutilization of
available resources. A study of the exact mechanisms through which
interference leads to unfairness reveals problems at multiple network
layers. The most general statements of these problems frequently
preclude finding a decentralized and optimal solution. However,
interference is a local disruption, and therefore leaves hope that a
local, if imperfect, solution may be found.

Distributed algorithms often make use of local exchanges of
information. This creates a need for a communication backplane capable
of connecting each host to the set of hosts affected by its
transmissions. This requirement is more cumbersome than it might seem
at first sight, for successful data transmission at common data rates
requires rather large signal to noise ratios. The capacity of links is
however affected even by interferers reaching them at far lower
signal levels. Connecting the recipients of interference with
transmitters requires thus a communication backplane capable of
operating over channels offering low signal-to-noise ratios.  This, in
turn, implies that constructing such a communication backplane will
require designing a physical layer different from the standard
physical layers used for high rate data transmissions.  Further
differences arise from the fact that, in a wireless environment
designed to support primarily data transmissions over short links
operating at strong signal levels, long range communication is at best
opportunistic.  Backplane communication receivers are therefore
required to be able to discriminate actual backplane transmissions
from high levels of background chatter.  Moreover, in order to offer a
significant improvement without further aggravating the interference
problem, communications along the backplane should not create new
interference constraints.

In this paper we propose a signaling scheme enabling the creation of a
communication backplane which meets all the above requirements. Our
scheme induces a low per-packet overhead, is resilient to high levels
of noise and interference, and minimizes the disruption of data
transmissions due to the interference that it induces.

Our first contribution is the design of tags, members of a set of
signals designed to be easily detectable and recognizable in the
presence of high levels of noise and interference, in the absence of
time and phase synchronization and with only approximate frequency
synchronization.
Their increase in range over regular data transmissions is obtained in
part through added redundancy. Tag signals are modulated using
multitone modulation over a time duration that is larger than the
duration of a regular data-transmitting tone. A tag is a distinct
superposition of several tones whose frequencies are chosen according
to the codewords of a binary algebraic code. On the receiver side,
tags are recognized using a receiver based on spectral analysis.

What about interference caused by tags on data payload?  OFDM, the
prevalent modulation for data transmission in today's wireless
networks, used in this study as well, exhibits flat time and frequency
power densities. 
We will prove that interference disruptions caused
by tags are comparable to the ones caused by data transmissions.

Our second contriboution is an analysis of the performance of tags at
different noise and interference levels and when making different
design choices. Starting from detection theory principles, we derive,
under a sufficiently general propagation model, the detection
likelihood/false alarm likelihood curves at different SNR levels. We
are particularly interested in quantizing the trade-off between
transmitting more information (i.e. more bits per tag) and the
corresponding increase in the likelihood of false alarm or
misclassification. Our findings are later on compared to experimental
results.

Our third contribution is the implementation and testing of Tag
Spotting through experiments performed using a software radio platform
in a testbed comprising senders, receivers and interferers.  This
series of experiments makes use of a tag family capable of conveying
about one byte of information.  The results of our evaluation support
the conclusion that communication through tags is effective at
SNR\footnote{Throughout this presentation we understand the noise part
  of the SNR figure to also include interference power, unless
  specifically noted otherwise.} values as low as 0 dB and is robust
to the effects of interference. In addition, through a series of
channel simulations, we establish the possibility of constructing tags
capable of conveying more bits of information and we survey the effects
of design decisions on the system performance.

Our fourth contribution is the use of Tag Spotting in two applications,
showcasing the performance improvements brought by the existence of a
control plane able to reach all nodes within the interference
range. Specifically, we use tags to efficiently implement a state of
the art congestion control scheme for multi-hop networks which
requires neighboring nodes, i.e, nodes that interfere with each other,
to exchange control information in an effort to fairly share the
available bandwidth. We also use tags in order to design and test a
simple MAC-layer signaling mechanism meant to prevent the starvation
of TCP flows in multi-hop wireless networks.

This paper is organized as follows: In Section
\ref{section:RelatedWork} we give an overview of related work and
indicate some congestion control and scheduling mechanisms that would
benefit from the use of tags.  Section \ref{section:AnOFDMPrimer}
introduces multitone modulation along with a simple model for
estimating the spectral footprint of multitone signals and discusses
techniques for limiting inter-carrier interference.  Armed with the
conclusions of Section \ref{section:AnOFDMPrimer}, we proceed in
Section \ref{section:TagSpotting} to describe in detail the structure
of tags and construct a tag detector capable of distinguishing tags
from interference. Section \ref{section:Structure} discusses in detail
the rationale behind each design decision present in tag construction.
Section \ref{section:Evaluation} experimentally evaluates, using a
software radio platform, the communication range of tags as well as
the rate of false detections. It is shown experimentally that tags can
be reliably identified at SNR values as low as 0 dB while the
likelihood of false detections can be sufficiently limited. The same
section evaluates the impact of noise, tag transmissions, and data
transmissions, on each other.  
Section \ref{section:Analysis} presents an analysis grounded on
detection theory principles of the effects of different choices
available in the process of designing the modulation of tags and, more
importantly, of the trade-off between the data rate of tags (i.e. the
number of bits transmitted) and the likelihood of false alarms.
Two examples of using tags to facilitate the implementation and
improve the performance of congestion control and scheduling are
presented in Sections \ref{section:CongestionControl} and
\ref{section:Scheduling}.  Finally, Section \ref{section:Conclusion}
concludes the paper.

\section{Related Work}
\label{section:RelatedWork}

{\bf Communication and Detection Theory.} Tag Spotting is closely
related to a classic research topic in communication theory, namely
information transmission at low signal-to-noise ratios. The motivation
of this research has varied from securing the transmitted data such as
in the case of spread-spectrum communication \cite{simon94spread} to
protection against interference in the case of the widely used CDMA
standard \cite{viterbi95principles} and to achieving long-range
transmission \cite{robin63multitone}.

\mindent Tags employ a multicarrier spread-spectrum modulation. They are
clearly related to MC-CDMA\cite{yee93multi}, however they use a
non-coherent modulation and do not use orthogonal codewords. Like
multitone FSK \cite{luo02achieving}, tags use a combination of tones
in order to transmit information.

\mindent The design of the tag detector presented in Section
\ref{section:TagSpotting} is based on the detection theory of
multipulse signals with constant amplitudes and unknown phases. While
the classical detector for such a situation is well-studied and
understood (see, for example \cite{mcdonough95detection,
  trees92detection}), it requires a precise estimate of the background
noise level in order to set appropriate detection
thresholds. Interference from competing packet transmissions will
confront tags with different levels of background noise, making a
precise and timely estimate impossible. Our detector
is independent of the level of background noise, requiring only a base
SINR as prerequisite for the accuracy of the detection. In the
appendix we apply a theoretical analysis similar to the one of the
classical detector in order to derive the detection/false alarm trade-off
curves of our own detector.

\vlad{Tags could in principle be constructed using well-established
  means of communication such as correlator-based detectors, standard
  spread-spectrum based modulations \cite{golomb04signal} or frequency
  shify keying.  However, these either assume a flat channel,
  preliminary synchronization in order to eliminate frequency and
  phase offsets or do not try to limit the maximal power in their
  spectral footprint (for the latter one).  The design of tags
  addresses all these concerns.}

{\bf Physical Layer Extensions.} In the wireless networking world,
carrier sense \cite{brodsky09indefense} can be seen as an example of a
message passing mechanism operating beyond the data transmission
range.  Closely related is the use of dual busy tones
\cite{haas02dual} in order to signal channel occupancy.  A recently
proposed physical layer extension, CSMA with collision notification
CSMA/CN \cite{sen10csmacn}, aims at reducing the impact of collisions
through an early termination signal sent by the receiver of the
colliding packet. The transmitter-based detector uses
self-interference cancellation techniques in order to improve the SNR
of the reciprocal channel and detects the termination signal using
correlation, in a manner similar to \cite{tufvesson99time}. However,
as the authors of both these papers find out, a correlation based
receiver cannot function without prior channel and frequency offset
estimation, which prevents their use for broadcasts over arbitrary
channels, as in the case of tags.

\mindent Carrier sense, dual busy tones and collision notifications
are binary signaling mechanisms, not suited for transmitting numeric
information, as required by message passing protocols.  Another recent
physical layer extension \cite{wu10side} aims at realizing a
side-channel over spread-spectrum based protocols through
perturbations of certain chips comprising a transmitted symbol. These
perturbations are in turn compensated for by the normal error
correcting codes employed in data transmission, thus allowing in most
cases for the payload to be decoded correctly, and they are also
detected by a special pattern analyzer, allowing fo rthe transmission
of side information. While the motivation and design constraints of
this work are similar to ours, the design choices made in the
creation of tags are very different. Most importantly, tags do not
require the data transmission scheme to be spread-spectrum based.

\mindent The technology of software defined radios \cite{usrp,
  gupta09clean} has acted as an enabler for some of the recent
advances in multiuser wireless network research.  It allowed, for
example, the experimentation of techniques such as zigzag decoding
\cite{golakotta08zigzag}, interference cancellation
\cite{halperin08taking} or dynamic bandwidth adaptation
\cite{chandra08case}.  
Perhaps the most similar technique to the one presented in this paper
is the one of smart broadcast acknowledgments, introduced in
\cite{dutta09smack}, in which multitone modulation is used for the
purpose of simultaneously conveying positive acknowledgments from
multiple receivers.
Other recent advances that also make use of software radios for
evaluation purposes include
multi-user beamforming \cite{aryafar10design}, fine grained channel
access in which bandwidth allocation based on an increased number of
carriers is coupled with frame synchronization used in creating an
effective uplink OFDM implementation \cite{tan10fine} and frame
synchronization used in obtaining diversity gains
\cite{rahul10sourcesync}. 

{\bf Congestion Control and Scheduling.} 
Prior works on congestion control for multi-hop wireless networks
differ in the way in which congestion is reported to the source.
One class of schemes sends implicit or imprecise feedback by dropping
or marking packets~\cite{xu03enhancing,rangwala08understanding} in the
tradition of TCP congestion control \cite{jacobson88congestion}, or by
regulating transmissions based on queue differentials~\cite{DiffQ}
along the lines of back-pressure ideas \cite{tassiulas92stability}.
Another class of
schemes~\cite{rangwala08understanding,tan07congestion} explicitly
computes available channel capacity and then sends precise rate
feedback, in the spirit of wired network congestion control
mechanisms such as XCP \cite{katabi02congestion} and RCP
\cite{dukkipati05processor}.

\mindent In an effort to tackle the complexity of creating optimal
schedulers, recent work on medium access for multihop networks has
proposed distributed algorithms capable of approximating the optimal
solution \cite{stoylar05maximizing} \cite{lin04joint}
\cite{jain03impact} \cite{kumar05algorithmic}
\cite{joo09understanding}. A common theme here is the use of local,
neighborhood-centered information in achieving a global solution.
Our work is partly motivated by the recent development of a number of
congestion control and scheduling schemes for multi-hop wireless
networks that are based on the local sharing of information, such as,
for example, \cite{rangwala08understanding}, \cite{chiang05balancing}
and \cite{chen06cross}.
Local information is at the heart of several other wireless
multi-hop problems: neighbor discovery \cite{vasudevan09neighbor},
reliable routing \cite{ye03aframework},
capacity estimation \cite{xu03enhancing} and signaling congestion and
starvation. The mechanism proposed in this paper offers an efficient
way to implement the neighborhood-wide sharing of control information
in the schemes mentioned.
\mindent While these schemes append control plane information to data packets
and rely on packet overhearing, it has been recognized that this
information needs to reach all nodes within the carrier sense range of
a node of interest.
The information sharing mechanism proposed in this paper eases the
implementation of many of these ideas and improves their performance
(see Section \ref{section:Applications}), as control information
will reach nodes outside the data transmission but within the carrier
sense range.

\section{An OFDM Primer}
\label{section:AnOFDMPrimer}

An OFDM encoder uses an inverse discrete Fourier transform in order to
encode a sequence of symbols into a set of tones over a finite time
interval, from here on named either a frame or an OFDM
symbol. Consider the sequence $ x $ of length $ N = 2^k $ whose
elements are chosen from a signal constellation, arriving for encoding
at an IFFT (inverse fast Fourier transform) frame generator. The resulting 
signal will be generated according to the formula:
\begin{equation*}
X(t) = \sum_{\forall k \in \{0, N-1\}} x_k e^{i k 2\pi t}.
\end{equation*}

\begin{figure}[tb]
\centering
\includegraphics[width=.35\textwidth]{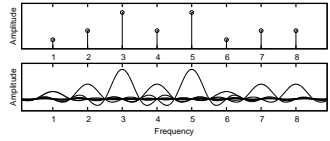}
\caption{The discrete and continuous Fourier transforms of an OFDM frame.}
\label{fig:ofdmprimer}
\end{figure}

\begin{figure*}[t]
\centering
\includegraphics[width=.95\textwidth]{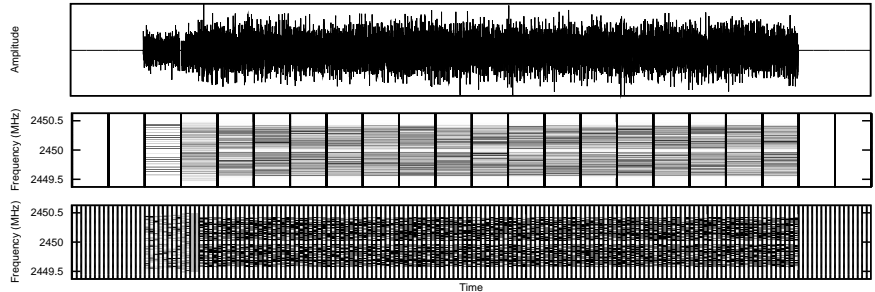}
\caption{Packet transmission: the signal $ s(t) $ (upper pane), a time
  frequency representation using 512 frequency bins (middle pane) and
  a time frequency representation using 64 frequency bins (lower
  pane).}
\label{fig:pktplot}
\end{figure*}

Passing to the continuous Fourier transform exposes a windowing effect
and provides us with an insight into the spectral footprint of the
generated signal. The discrete spectrum of an OFDM frame is
illustrated in the upper half of Figure \ref{fig:ofdmprimer} and
corresponds to the original encoded symbol sequence $ x $. Since the
summed exponentials are bounded in time, their continuous transforms
are sinc functions and the spectrum of the encoded signal is the sum
of these sinc functions, as depicted in the lower half of Figure
\ref{fig:ofdmprimer}. The orthogonality of different signals is
preserved: the peak of each sinc function is aligned with the zeros of
all other sinc functions.

Let's focus on non-coherent communication, i.e. we assume that the
phase of each received signal is independently random, and try to
derive a concentration result for the power leaked outside the
intended transmission bandwidth.  In Section \ref{section:TagSpotting}
we will use this result in order to motivate that the modulation of
tag signals exhibits a fast spectral decay and limited inter-carrier
interference. The reasoning behind our approach is encompassed in
Equations (\ref{eq:leakage_equation}) and (\ref{eq:euler_equation})
below, which describe the effects of self-interference on OFDM signals
and give the speed of their spectral decay. In the presence of a small
frequency offset $ \delta $, the interference signal added by a
carrier with unitary amplitude to a carrier placed $ k $ positions
away is
\begin{equation*}
l(k, \delta) \, = \, \int_{[0,1]} e^{2\pi i (k+\delta)t} dt \, = \,
\frac{\sin(\pi \delta)}{\pi(k+\delta)}e^{i\pi\delta},
\end{equation*}
and in power terms,
\begin{equation}
p(k, \delta) \, = \, \mathrm{sinc}^2(\pi(k+\delta)) \, \leq \,
\frac{1}{(k+\delta)^2}.
\label{eq:leakage_equation}
\end{equation}

It can be easily verified that the magnitude of the interference among
two carriers does not depend on the absolute difference of carrier
frequencies but only on the number of carrier positions separating
them. 

Consider now a carrier at carrier position zero and an infinite block of
carriers starting at carrier position $ k $. The average power leaked
by this block of carriers into the zero-positioned carrier can be
bounded according to Equation \eqref{eq:leakage_equation} to be:
\begin{equation}
p_{b}(k) \, = \, \sum_{c=k}^{\infty} \frac{1}{c^2} \, = \,
\frac{\pi^2}{6} - \sum_{c = 1}^{k-1} \frac{1}{c^2},
\label{eq:euler_equation}
\end{equation}
where we assume that symbols on different carriers are
independent. The series in Equation (\ref{eq:euler_equation}) is
rapidly converging. 

Based on the observations above we conclude that replacing a single
carrier through a block of random-phase, lower-bandwidth carriers that
occupies the same part of the frequency band will sharpen the spectrum
of a multitone signal and accelerate its spectral decay.  Moreover,
the insertion of a relatively small number of null-carriers will
significantly reduce interference between two neighboring active
regions of the spectrum.  Creating the replacement carrier block calls
for an increase in the number of carriers, and a corresponding
increase in the time length of the OFDM frame, as described fully in
Section \ref{section:TagSpotting}.

\section{Tag Spotting}
\label{section:TagSpotting}

{\bf High Level Design Overview.} 
   Tag Spotting uses a set of signals (tags) that are easily
   detectable in low SNR conditions. The design of tags is determined
   by a number of constraints.  Firstly, due to their short timespan,
   the presence of multiple indistinguishable sources and the presence
   of varying levels of interference, the tag detector cannot perform
   accurate channel estimation, or achieve time, phase and fine
   frequency synchronization. Secondly, in order to protect competing
   data transmissions from further levels of interference, tags must
   abide by a maximal spectral power constraint, which prevents the
   use of a peaky transmission scheme such as multiple frequency shift
   keying (MFSK). Finally, a tag detector must perform identically in
   the presence of added interference, as long as the SINR remains
   unchanged.

\mindent To address these constraints, we employed a noncoherent
communication scheme that spreads a tag's energy over the frequency
domain. \vlad{In constructing tags, }the number of carriers was
increased \vlad{from the one used in payload data transmission} in
order to sharpen the signal's spectral footprint and ease spectral
analysis through the discrete Fourier transform even in the absence of
complete frequency synchronization. The symbol sequences encoded over
the different carriers were chosen according to an algebraic code,
thus adding extra redundancy.

The current section presents the details of tag construction. The
following section will present the reasoning behind the design decisions
made throughout the construction.

{\bf Multitone Structure. }  Figure \ref{fig:pktplot} presents a
  packet transmission in the time domain representation as well as in two
  different time-frequency representations. The two time-frequency
  representations, pictured in the lower two panes, are realized by
  performing the spectral analysis of successive blocks of 512 samples
  (middle pane) or 64 samples (lower pane). As we will see, these
  lengths are natural choices for describing the structures of tags
  and packet payloads respectively. In the same representations, each
  vertical column corresponds to an analyzed block, while the
  horizontal line pattern present in each such column illustrates the
  block's power spectral decomposition. In order to make the
  representation more meaningful, we have stripped tags and data
  frames of their cyclic prefixes and we aligned the boundaries of the
  analysis periods with the boundaries of tags and data frames.

\begin{figure}[tb]
\centering
\includegraphics[width=.4\textwidth]{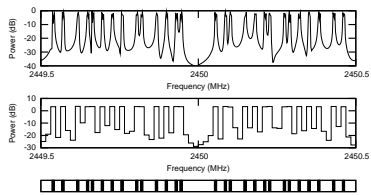}
\caption{The spectrum of a received tag in the presence of a frequency
  offset. Upper pane: the 512 frequency bins Fourier transform of the
  corresponding detector interval. Middle pane: the 64 frequency bins
  representation used in the detection decision. Lower pane: the
  structure of the transmitted tag.}
\label{fig:tags}
\end{figure}

\mindent The structure of a tag in frequency domain is depicted in the
lower pane of Figure \ref{fig:tags}. Tags are encoded using 512 OFDM
carriers while packet information makes use of 64 wider data
carriers. Therefore, in time domain, a tag spans a period equivalent
to eight regular data frames, while in frequency domain each ``wide''
data carrier corresponds to eight ``thin'' tag carriers. The
transmission begins with a tag, whose spectrum can be observed in the
third column of the middle pane, and continues with data frames
containing synchronization data followed by the packet's payload. To
both tags and data frames we append a cyclic prefix which increases
their respective lengths by $\frac{1}{4}$-th.

\begin{figure}[tb]
\centering
\includegraphics[width=.4\textwidth]{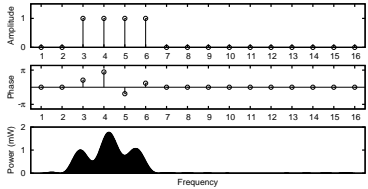}
\caption{The discrete spectrum of a two-carrier group (amplitude and
  phase representation) and its continuous power spectrum (bottom panel).}
\label{fig:header}
\end{figure}

\mindent The spectrum of a tag is constructed, at a basic level, by
transmitting a symbol from a 0-1 (on/off) constellation in the
frequency space corresponding to each of the wide carriers. Wide
carriers are in turn grouped into groups of two neighboring carriers,
and in each group only one of the two wide carriers will be
active\vlad{, carrying the entire power transmitted through the
  group}. This last constraint renders the modulation inside each
carrier group to be a form of binary frequency shift keying.
Such a two-carrier group is depicted in Figure \ref{fig:header}.
On the receiver side, the tag detector
operates in the presence of small, tolerable but unknown frequency
offsets, which may cause power spillage from the active wide carriers
to the inactive carriers.
As illustrated in Figure \ref{fig:header}, we chose to send the entire
signal power allotted to an active wide carrier using the central four
of the eight thin carriers corresponding to this particular wide
carrier. We also chose to encode the tones sent on these four thin
carriers using symbols of the same amplitude but which use
\vlad{uniformly} independent random phases. Looking at Equation
\eqref{eq:euler_equation} \vlad{, since the use of the random phases
  implies that the symbols transmitted over the different thing
  carriers are independent}, it results that \vlad{increasing the
  number of carriers and using random phases for the transmitted
  symbols } reduce the amount of power leaked onto inactive wide
carriers.\footnote{\vlad{More precisely, Equation
    \eqref{eq:euler_equation} predicts that under these circumstances
    the signal's roll-off, considered anywhere in the frequency band,
    is compressed through a factor that equals the increase in number
    of carriers when replacing wide carriers with thin carriers.}}

A straightforward computation assuming a frequency offset
distributed uniformly between zero and two thin carrier widths reveals
that the expected power leaked is, in expectation, about 2.3\% of the
total power. Figure \ref{fig:tags} further illustrates this aspect by
presenting the spectrum of a received tag in the presence of a
frequency offset equal to 10\% of a regular carrier width. By
comparing the distribution of the received signal power in the
frequency bins corresponding to wide carriers (middle pane) with the
structure of the transmitted tag (lower pane), it can be seen that the
received power is concentrated in those bins which correspond to
active wide carriers.  As a note, the use of random phases in signal
construction has one further advantage: it allows sampling tags from a
larger signal set in order to limit their peak to average power ratio.

\vlad{For tag construction purposes, eight of the wide carriers have been
designated null carriers, while the remaining 56 carriers give rise to
28 two-carrier groups. Every tag can thus be naturally mapped to a
28-bit string in which each bit marks the choice of state in one of
the groups.}

{\bf Encoding. } As it is common in communication system design,
  modulation is supplemented by a coding layer. The purpose of this
  layer is to create a subset of maximally differentiable signals,
  and to reduce the number of hypotheses tested.

 The tag signal construction detailed above produces $ 2^{28}
$ different basic tags, too many for efficient detection and
insufficiently distinguishable from each other. A further restriction
makes use of a $(28, 60, 13)$ \footnote{The notation for codes used
  here is in the form $(N,M,D)$ where $N$ is the binary codeword
  length, $M$ denotes the number of distinct codewords and $D$ denotes
  the minimal Hamming distance between any two codewords.}  nonlinear
code proposed by Sloane and Seidel \cite{sloane70nonlinear}. Out of
the basic tags only those having binary representations corresponding
to the 60 codewords of this code are preserved.
We discuss the use of other codes in Section \ref{section:Evaluation}.

 Regular data frames spread the transmitted power over 48
carriers while tags make use of only 28 carriers. Since the average
power spectral densities of used carriers in tags and data
transmissions are equal, it results that the transmitted power in the
case of tags is lower than the transmitted power in the case of data,
as it can be observed in Figure \ref{fig:pktplot}.

{\bf Constructing a Tag Detector. } Let $ \mathit{T} = \{t_1,
  t_2, \ldots, t_{60}\}$ denote the set of all tags and $ C_i $ be the
  set of all data carriers activated when transmitting tag $ t_i $.
  Let $ r_f $ denote the power of the received signal in the frequency
  bin corresponding to the $f$-th carrier. Our detector does not
  assume the channel phase response to be uniform and can therefore be
  used in a wideband scenario. We compute the following quantity which
  we will name from now on tag strength:
\begin{equation}
\label{equation:tag}
\frac{\sum_{{f \in C_i}} r_f^2}{\sum_{\forall f} r_f^2}.
\end{equation}

 Tag strength is compared against a fixed threshold $ \gamma $
and in case the threshold is exceeded a possible tag observation is
recorded. \footnote{This equation is similar to the one of the low-SNR
  multipulse detector with $n$ samples for a signal with unknown phase
  $ \theta(t) $ varying at each pulse: $ s(t) = A*\cos(2 \pi f
  t+\theta(t)) $, at a given SNR value $ \alpha = \frac{A}{2N}
  $. Denote by $ H_S $ the hypothesis that signal $ s $ has been sent
  and by $ H_n $ the hypothesis that no signal has been sent. That
  detector is based on the equation $\log{\frac{p(r|H_S)}{p(r|H_n)}}
  \approx \sum_{i=1}^{N} r(t_i)^2 > \gamma' \frac{N}{\alpha} $ where $
  \gamma' $ is a constant (see \cite[p. 293]{mcdonough95detection});
  in our case, the correction factor $ \sum_{\forall f} r_f^2 $ can be
  seen as an approximation of $ \frac{A^2}{4}+N = (\alpha+1) N$, meant
  to remove the linear dependence of the threshold on the noise power
  $ N $, allowing thus for added noise-like interference.}

 Detection intervals have the same length as a tag from which
the cyclic prefix has been removed and are spaced one tag cyclic
prefix length apart. It results that successive analyzed intervals
have significant overlap. Every transmitted tag will completely cover
at least one detection interval. The detector processes every interval
by first computing the Fourier transform of the contained signal and
then computing, based on the resulting spectrum, the strength of each
tag according to Equation \eqref{equation:tag}. In order for a tag
recognition event to be recorded, the corresponding tag strength must,
firstly, exceed the threshold value and, secondly, be maximal among
all tag strengths (for all tags) derived from detection intervals that
overlap the current interval.  A further detection metric effective in
filtering off-band interference is computed for each detection
interval by weighting the power levels in different carriers through
the carrier's position in the frequency band, summing the resulting
values and afterward dividing the result through the total interval
power. As long as the resulting ``center of mass'' is placed in the
central quarter of the frequency band, the tag observations are
considered valid, otherwise they will be attributed to off-band
interference.

 In order to reduce the number of intervals analyzed and the
likelihood of false alarms, a simple carrier sense scheme is
employed. The receiver maintains a running estimate of channel noise
and processes only those intervals for which the SNR exceeds $ -1 $
dB.

{\bf Overhead.}  Adding a tag to a packet incurs a transmission
  time overhead. Assume for now that only data packets are tagged and
  that a typical data packet has a payload of about 1500
  bytes. Encoded using the parameters presented in Section
  \ref{section:Evaluation}, the payload will span 125 data frames, to
  which six synchronization frames are appended. A tag spans the
  equivalent of eight data frames and therefore its overhead is about
  6.1\% in terms of the normal packet duration. In some control
  schemes some of the data messages will not require tags to be
  piggybacked, allowing for a lesser overhead.

\section{Motivating the Design Choices}
\label{section:Structure}

The previous section has presented in considerable detail the
structure of tags. While the above description is complete, the
decisions taken in the construction of tags may well seem
arbitrary. The purpose of the current section is to motivate every
design decision taken in tag construction.

\subsection{Multitone Structure}

The design space of tag signals is frequency space. At the lowest
level, tags use the same form of modulation in frequency space, using
orthogonal signals over a finite time interval, that is used in
regular data transmission.

{\bf OFDM.}  
  In an opportunistic reception system that searches for the short
  occurrences of tags, the price of exact frequency synchronization
  and of exact channel estimation should be avoided. Tags are
  therefore transmitted and received without performing channel
  estimation or frequency synchronization.  This raises a challenge in
  solving inter-carrier interference.

\mindent Remember that the choice of OFDM for data transmission is
linked to the propagation behavior of wireless signals. Since sine
signals are the eigenfunctions of the wireless channel, the use of
orthogonal sines along with an appropriate cyclic prefix is meant to
prevent any channel-caused interference among the different
transmitted symbols.
  The limited timespan of the transmission interval precludes the use
  of actual sines or any other signals with narrow support in the
  frequency space.  As it was discussed in Section
  \ref{section:AnOFDMPrimer}, the actual signals used in the
  transmissions have a slowly-decaying spectral footprint.  In OFDM
  data transmission, the inter-carrier interference which the slow
  spectral decay entails is avoided through exact frequency
  synchronization, using the fact that in frequency space the zeros of
  the base sine-like signals align with the peaks of all other signals
  in the base set.
  In contrast, for tags, the packing of noncoherent carriers into
  larger building blocks reduces the amount of power leaked among
  the frequency bins corresponding to thick carriers. This reduction
  in leaked power allows the system to function as intended even when
  the receiver is not frequency-locked onto the transmitter.
  The lack of frequency synchronization together with the lack of an
  estimate of the channel phase response at different frequencies also
  leads to uncertainty regarding the phase of the transmitted
  signal.  The use of a noncoherent encoding
  allows us to overcome the lack of knowledge of the channel phase
  response without further complications.  
  It could be argued that these channel characteristics should be
  measured in advance. However, our receptions are at best
  opportunistic, and the channel could be any one of a multitude of
  fast changing channels between any pair of hosts. Certainly,
  obtaining an exact estimate of the channel response and of the
  frequency offset, at the low SNR levels for which our system is
  designed, would greatly complicate the tag transmission problem.

 {\bf Fading.} Another characteristic of wireless channel
  transmission, frequency-selective fading, provides the rationale for
  the use of groups of two carriers as a encoding unit: since
  neighboring data carriers are likely to experience similar fading
  and since any of the codewords makes use of either one or the other
  of the two carriers in a two-carrier group in order to encode a bit
  value, fading over a two-carrier group will not induce a bias
  towards any of the hypotheses that a particular codeword has been
  transmitted. The received power and the likelihood of detection may
  well decrease due to fading. However, when considering a given
  overall signal to noise ratio, i.e. computed over all the carriers,
  the detection probabilities over fading and non-fading channels are
  quite similar, as shown in Section \ref{section:Analysis}, while the
  false alarm probabilities are the same. We can conclude that this
  particular design decision manages to overcome most of the
  difficulties that fading introduces in tag detection.

\subsection{Constructing a Tag Detector}

{} \mindent It is worth mentioning here a significant difference
  between the main purpose of a tag receiver and the purpose of a
  communication system receiver. While a communication system receiver
  is meant to accurately distinguish between a number of hypotheses
  corresponding to signal transmissions under the assumption that an
  actual transmission has occurred, the tag receiver listens for the
  most part to noise and background chatter. The main task of a tag
  receiver is therefore to detect, with sufficient confidence, a tag
  transmission when one occurs and, if possible, to correctly identify
  the transmitted tag. Tag detection is therefore a detection problem
  more than a communication problem and the design of the detector
  reflects this fact. The probabilities of false alarm allowed in the
  case of tags are well under the typical probabilities of
  misclassification allowed in a communication system, since the
  occurrence of tag transmissions are assumed a priori to be rather
  rare events. False alarms weigh in more heavily when
  compared to the total number of detected tags.  Due to the fact that
  tag detection is essentially a detection theory problem, we choose
  the detection metric (probability of false alarm versus the
  probability of detection) to be the main measure of tags
  performance. The secondary metric considered will be the probability
  of misclassification of a transmitted tag. The experimental section
  will reveal that this probability is negligible due to the high
  threshold required for a positive tag detection, even when using a
  rather large number of codewords.

 {\bf Detecting Patterns.}  In general, tag observations occur
  over short intervals of time and channel conditions change too
  frequently for the receiver to obtain and update an accurate noise
  and interference power estimate. The only assumption made in the
  following is that the spectral envelope of the noise and
  interference signals is flat, an assumption that can be justified in
  the case of data networks using OFDM-based encoding. We design
  therefore our modulation scheme and our receiver to use as a
  detection indication not the sheer amount of power received but
  rather the concentration of the received power into pre-determined
  frequency bins. The receiver detects a transmission event whenever
  the concentration of the received power (the ratio of the power
  received in the designed frequency bins to the overall received
  power) exceeds a certain threshold. Therefore, the receiver searches
  not just for the presence of a signal but for a certain spectral
  shape. The fact that a power ratio measurement is used as a
  detection metric guarantees an universal receiver in a wider sense:
  the probability of detection for a threshold value chosen as the
  receiver parameter will only increase with increasing SINR. A
  standard detector is denoted as universal when a similar guarantee
  exists in terms of the SNR.

 {\bf Tags and FSK.} The attentive reader might have noticed that
  a simpler encoding scheme might have provided a similar
  detection/false alarm performance trade-off without the use of an
  algebraic code. Frequency shift keying simply concentrates the
  available power into the frequency space corresponding to one of the
  available carriers, thus offering similar received power
  characteristics. However, FSK has a large power spectral density,
  due to the fact that all the transmitted power is effectively
  concentrated in one point of the frequency spectrum, which makes it
  undesirable in a network environment, where we would like to
  guarantee a certain flat envelope for the frequency spectrum of our
  transmission, with a fast decay outside the data band. Our choice of
  modulation limits the transmitted power at any given frequency,
  resulting thus in a flat spectrum, similar to the one corresponding
  to OFDM data transmissions. The experiments in Section
  \ref{section:Evaluation} verify that the typical interference
  effects of tags on competing data transmissions are not worse than
  the interference effects caused by normal packet data transmissions
  sent at a similar overall power level, which would not be the case
  if tags were modulated using FSK.

 {\bf Choosing the number of active carriers.} In Section
  \ref{section:TagSpotting} it was mentioned that tags are a specific
  form of multiple frequency shift keying (MFSK), a modulation which
  makes use of multiple noncoherent carriers transmitted simultaneously
  as a single symbol. A key parameter of MFSK is the number of
  transmitted carriers. Having chosen the MFSK modulation and the
  power ratio detector as the basic building blocks of our system, we
  must find next a value of this parameter that offers reasonable
  detection performance while also allowing for the construction of a
  large set of tags. As a definition of performance, we seek to
  minimize the probability of false alarms while preserving a certain
  probability of detection.
  In Section \ref{section:Analysis} this measure of detection
  performance is evaluated, as a function of the number of carriers
  transmitted, in the case of a detector that is searching for a
  single tag. It is revealed that, for a detector operating at a
  target SNR of 0 dB, the optimal allocation of power uses a bit less
  than half of the available carriers. Therefore, the decision to use
  exactly half of the carriers in the modulation does not impact the
  detector's performance, while the same decision eases the use of an
  algebraic code. 

\subsection{Encoding}

 {\bf Families of Codewords.} The decision to use
  exactly half of the carriers in the construction of tags is
  motivated by the details of constructing a tag family. In
  particular, we look at the relationship between tags and the binary
  codes used in their construction, namely codes over the same number
  of bits as the number of thick carriers. The simple binomial
  expansion indicates that the binary strings whose weight is half
  this number of bits are most numerous. We expect that for small
  Hamming distances the codes of this particular fixed weight would
  have more members than codes constructed using any set of binary
  strings of a different fixed weight. Our construction, which
  converts codeword families constructed over a number of bits
  equaling half the number of thick carriers into families of
  codewords of fixed weight over twice that length, is almost optimal,
  as it can be checked using tables of optimal known codes of fixed
  weight. Moreover, this construction allows for the use of well-known
  algebraic codes in tag construction and therefore leaves open the
  possibility of developing fast algorithms for identifying the most
  likely codeword in the case of large codeword families, in a manner
  similar to the identification, in communication systems, of the most
  likely transmitted codeword based on receiver soft symbol values.

 {\bf Construction Procedure.}  Based on this description of the
  intricacies of tag construction we are now able to give a general
  procedure for constructing a set of tags that meet a desired set of
  performance criteria. Firstly, a codeword family that guarantees a
  low enough probability of misclassification at the target SNR while
  offering a sufficient number of different codewords must be
  selected. The probability of misclassification is computed in this
  step using Monte Carlo simulation, as presented in Section
  \ref{section:Analysis}. Secondly, again using Monte Carlo
  simulation, as presented in Section \ref{section:Evaluation}, the
  tag designer must determine the probabilities of false alarm for the
  chosen tag family. The probability of detection can be more readily
  computed since it does not depend on the choice of codeword family
  but only on the number of carriers that a codeword uses. This
  process must be iterated until a suitable trade-off between the
  number of codewords available, the probability of detection and the
  probability of false alarm is achieved. When satisfactory results
  cannot be achieved, the tag designer may increase the time length of
  tag through the use of a larger FFT window and a correspondingly
  increased number of thin carriers, adding more redundancy to the
  signals.

\section{Evaluation}
\label{section:Evaluation}

\subsection{Experimental Setup}

{\bf (a) System Parameters} The experiments were conducted using two
USRP boards \cite{usrp}, one transmitter and one receiver, in an
indoor environment without line of sight and with multiple concrete
walls between the sender and the receiver. The GNU Radio software
suite was used for signal transmission and recording.  The carrier
frequency chosen was 2.45 GHz and the bandwidth of the system was set
to 1 MHz.
In order to obtain a linear channel suitable for OFDM transmission
without distortion effects introduced by the transmitter mixer and the
receiver, the signal has been oversampled by a factor of four on both
transmitter and receiver sides.  
The system uses the Schmidl and Cox algorithm \cite{schmidl97robust}
for packet detection and initial CFO estimation and the Tufvesson
\cite{tufvesson99time} algorithm for block boundary detection.
Channel gain is measured using the preamble, while phase response
is tracked using four pilot carriers and a linear phase interpolator.
The symbol constellation used was 16-QAM. Packet payload was encoded
using a rate $1/2$ punctured Trellis code. The resulting link speed is
2.4Mbps.

\mindent While the design of an efficient OFDM transmission and
reception system is not the focus of this presentation, having a
viable system was a precondition for showing that any interference
effects produced by our set of header tags are comparable to
interference effects caused by data transmissions and by environmental
noise.

 {\bf (b) Experiment Description.} We performed three series of
  experiments intended to evaluate the impact of decreasing the signal
  to noise ratio on the effectiveness of tag spotting, the impact of
  rising interference power on tag spotting and the disruption caused
  to data transmission by interference in the form of tags. 
In order to determine the likelihood of false alarms, we have
conducted a further series of experiments using half minute-long
recorded signal sequences containing ambient radio noise pertaining to
standard 802.11b/g transmissions in an office building occupied by
numerous wireless networks in order to measure the detector's
robustness to different kinds of radio interference. We have also
evaluated through simulations the likelihood of misclassifications.

\begin{figure*}[t]
\centering

\subfloat[Experimental results in the presence of channel noise.]
{
\includegraphics[width=.4\textwidth]{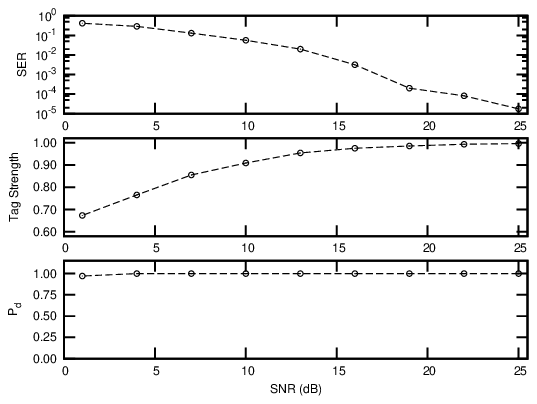}
\label{fig:expnonoise}
}
\hfil
\subfloat[Experimental results in the presence of data-like interference.]
{
\includegraphics[width=.4\textwidth]{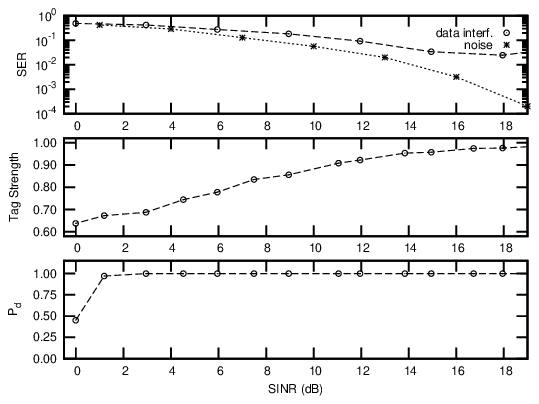}
\label{fig:exptagpktnoise}
}

\subfloat[Experimental results in the presence of tag-like interference.]
{
\includegraphics[width=.4\textwidth]{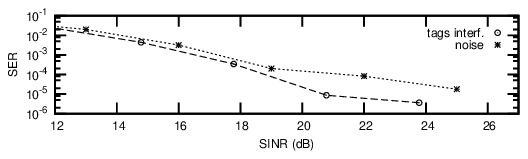}
\label{fig:exptagsignoise}
}
\hfil
\subfloat[Average time between false alarms at different thresholds.]
{
\includegraphics[width=.4\textwidth]{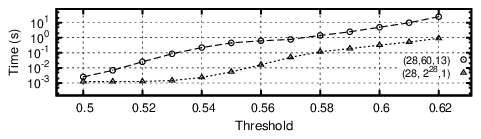}
\label{fig:exppfa_t}
}
\caption{Experimental Results.}
\end{figure*}

 {\bf (c) Metrics}

\begin{description}

\item[Tag Strength] is the quantity defined in Equation
  \eqref{equation:tag}, the primary metric for deciding whether a tag
  observation will be recorded. It is a measure of the ratio of power
  contained in the frequency bins allotted to a given tag and the
  total received power. In order for a tag observation to be recorded
  one of the necessary conditions is that the tag strength must exceed
  a threshold value $ \gamma $. In all experiments presented $ \gamma
  = 0.62 $ was used, \vlad{which was chosen using the theoretical
  performance curves derived in Section \ref{section:Analysis}}. This
  choice of threshold accomplishes two goals: it is high enough to
  correspond to a low rate of false alarms, as verified through the
  experimental results presented in the current section (see Figure
  \ref{fig:exppfa_t}) and it is low enough to allow detection at the
  target SNR values. Using the na\"{\i}ve assumption that noise (and
  interference) power contribute equally to the power levels detected
  in the different frequency bins, the SNR value that corresponds to
  this threshold can be derived to be about 0.4 dB.

\item[Symbol Error Rate (SER)] is measured for the payload of all
  correctly identified packets, that is, packets for which the packet
  detection, block boundary start estimation and CFO estimation
  succeed. It is the primary metric for estimating the effects of
  various noise and interference levels on data transmission. This
  metric was considered more fundamental than the bit error rate(BER),
  which is heavily dependent on the type of coding employed, a system
  design parameter that varies largely in current designs.

\item[Probability of Detection ($P_d$)] is defined as the probability
  that a header tag will be correctly detected and identified at
  different SNR and SINR levels. It is the primary metric for the
  success of tag spotting.

\item[Probability of False Alarm ($P_f$)] is defined as the likelihood
  that, in any given detection interval, noise and interference will
  cause a spurious tag detection and identification in the absence of
  a tag transmission.

\item[Probability of Misclassification ($P_m$)] is the likelihood of
  incorrect tag identification in the presence of a tag transmission.
  
\end{description}

{\bf (d) Practical Considerations.}  The tag detector presented in
Section \ref{section:TagSpotting} is unable to compensate for
frequency variations between the sender and the receiver. The
structure of tags and the detection method described makes it possible
to tolerate frequency offsets of up to two thin carriers, or about 4
kHz, without perceivable performance impacts. We have found that the
clock jitters of the USRP radios are well within this limit, however
different radios have initial frequency offsets of up to 200 kHz,
necessitating a supplementary calibration step before each
experiment. We assume the effects of Doppler spread to be minimal,
i.e. a near-static scenario.

In a practical scenario the clock components could be replaced with
more accurately designed/packed counterparts, and therefore we
conclude that constructing self-sufficient tag detectors is possible.
\vlad{Increasing the baseband bandwidth from the relatively narrow
  bandwidth of the USRPs (1 MHz for OFDM experiments) to the standard
  20-40 MHz of WiFi would enlarge the thin carriers and
  correspondingly allow for larger frequency offsets which would make
  tags even more robust.}

\subsection{Experimental Results}

{\bf (a) Impact of Noise.} The first series of experiments tries
  to quantify the range effectiveness of tag spotting in the presence
  of different levels of noise, in an interference-free environment.

\begin{figure}[tb]
\centering \includegraphics[width=.36\textwidth]{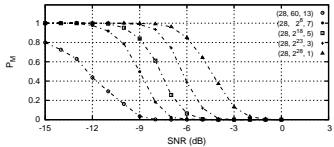}
\caption{Probability of tag misclassification at different SNR levels.}
\label{fig:exppm}
\end{figure}

\mindent The transmitter was configured to send sequences of 100 packets with
random header tags. On the receiver side the transmitted sequence was
decoded and the sequence of detected tags was compared to the original
transmitted sequence, in order to obtain an estimate of the detection
probability $ P_d $. The decoded symbol payload of received packets
was compared with the known symbol payload on the transmitter side in
order to estimate the SER . The transmission's SNR was estimated for
each detected packet using a low-pass filter-based average power
estimator.
The power level of the transmitter was varied between levels spaced 3
dB apart, resulting in different channel SNR values.

\mindent Figure \ref{fig:expnonoise} illustrates the results obtained.
The upper pane shows the Symbol Error Rate (SER) for the payload as
a function of the Signal to Noise ratio (SNR). The curve is typical
for a receiver employing 16-QAM modulation, however the receiver
appears to exhibit an error floor at the higher SNR values
measured. At SNR values of 20-25 dB, the system can sustain data
transmission, when using a typical error-correcting code. This curve
serves as a reference for the next experiments, in which noise-based
disruptions will be replaced with data-like interference and tag-like
interference.

\mindent The middle pane shows the Tag Strength as a function of the
SNR. The curve decreases steadily as the SNR decreases, reaching the
threshold value $\gamma$ around 0 dB.

\mindent Finally, the lower pane shows the probability of detection as
a function of the SNR. It can be seen that the probability of
detection is close to one over the entire range considered.

{\bf (b) Impact of Interference.} Figure \ref{fig:exptagpktnoise}
  present the results of the same experiment in the presence of a
  second source transmitting an uninterrupted stream of payload-like
  data.  As before the upper pane plots the Symbol Error Rate, the
  middle pane the Tag Strength, and the lower pane the probability of
  detection, all as a function of the SNR.  The SER has a slightly
  different behavior in this case, due to the presence of a different
  type of interference, as can be seen when comparing the SER curve in
  the presence of data interference with the SER curve in the presence
  of just noise. The other quantities of interest, tag strength and
  the probability of detection $ P_d $ remain essentially
  unchanged. The probability of detection climbs a steep curve and
  quickly settles close to one. We conclude that the tag detector acts
  almost identically in the presence of pure noise or noise combined
  with temporary interference.

{\bf (c) Impact of Tag Interference on Data.} Figure
  \ref{fig:exptagsignoise} presents the effect of tags on data
  transmissions. The SER curve is very close to the SER curve of
  Figure \ref{fig:expnonoise}, demonstrating that interference from
  tags does not increase the error likelihood beyond the error
  likelihood in the presence of comparable levels of noise.

{\bf (d) Likelihood of False Alarm.} Figure \ref{fig:exppfa_t}
  presents the dependence of the average time in-between false alarms
  on the threshold $\gamma$, when analyzing recordings of ambient WiFi
  traffic. Carrier sense has been disabled in this experiment and
  every input detection interval is analyzed. These results support
  our choice of detection threshold, since false alarms occur at a
  rate of less than once every 20 seconds.

{\bf (e) Likelihood of Misclassification.} Figure \ref{fig:exppm}
  presents the dependence of the probability of misclassification on
  the receive SNR when tags are constructed using either the code
  mentioned in Section \ref{section:TagSpotting}, a few extended BCH
  codes with smaller minimal distance or an unencoded modulation. No
  lower tag strength threshold was employed. This plot reveals that,
  for all these schemes, misclassification does not occur at the
  targeted signal levels. More details on the significance of this
  result will be given in the next section.

 {\bf (f) Choice of algebraic code.} 
The choice of algebraic code affects two quantities of interest, $P_m$
and $P_f$. It has already been noted that, for a large class of codes,
the probability of misclassification is negligible at the targeted
signal levels. On the other hand, the probability of false alarm will
increase as the number of codewords increases, as illustrated in
Figure \ref{fig:exppfa_t}, due to the presence of supplemental
hypotheses. It results that there is a trade-off between the number of
bits of information available per tag and the desired rate of false
alarms.

\subsection{Tag Range}

The possible use of tags in wireless multi-hop networks prompts us to
calculate the increase in range, expressed in distance terms, that
transmission through tags brings over regular data transmission.  It
is well-known that power decay exponents are highly dependent on the
environment. Measurements of signal propagation in the 2.4 GHz band
described in \cite{paul08wireless} reveal a dependence of received
power on distance of the form $ p \propto \frac{1}{r^d} $ where the
exponent $ d $ varies from 3 for line of sight propagation to about 6
for non-line of sight propagation. Regular data transmissions using
the 16-QAM modulation present in our system necessitate a SNR value of
about 20 dB while tags are detectable at a SNR value of 0 dB. It
results that the ratio of the range of tag communication to the range
of the data transmissions can be, depending on the power decay
coefficient, anywhere between 2.15 and 4.65.

The carrier sense threshold is usually set about 10-20 dB above the
noise level \cite{brodsky09indefense} \cite{vasan05echos}, which
substantiates our claim that Tag Spotting can communicate information
beyond the carrier sense range.

\section{Performance Analysis}
\label{section:Analysis}

\begin{figure*}[t]
\centering
\subfloat[one-tag (wideband model)]
{
  \includegraphics[width=.3\textwidth]{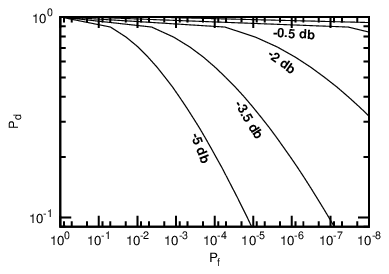}
  \label{fig:analysis_1tag}
}
\hfil
\subfloat[the $(28, 13, 60)$ code (wideband model).]
{
\includegraphics[width=.3\textwidth]{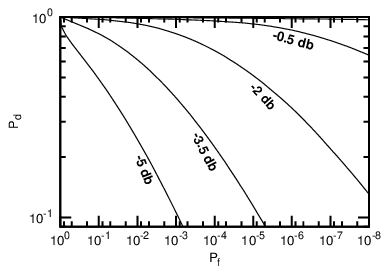}
\label{fig:analysis_60tag}
}
\hfil
\subfloat[the $(28, 1, 2^{28})$ code (wideband model).]
{
\includegraphics[width=.3\textwidth]{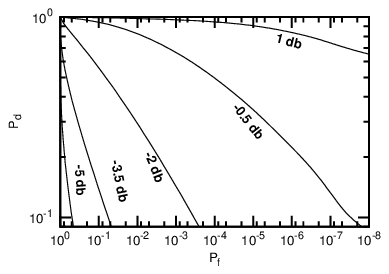}
\label{fig:analysis_alltag}
}
\\
\subfloat[one-tag (narrowband model)]
{
\includegraphics[width=.3\textwidth]{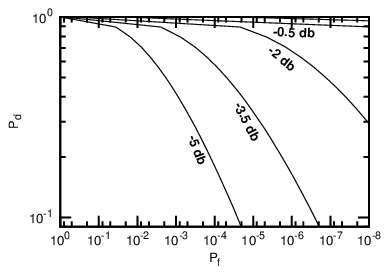}
\label{fig:analysis_1tagexp}
}
\hfil
\subfloat[the $(28, 13, 60)$ code (narrowband model).]
{
\includegraphics[width=.3\textwidth]{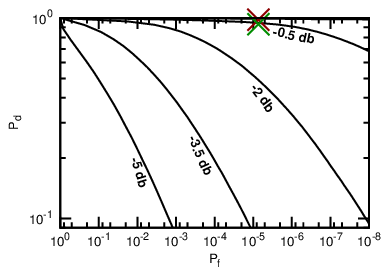}
\label{fig:analysis_60tagexp}
}
\hfil
\subfloat[the $(28, 1, 2^{28})$ code (narrowband model).]
{
\includegraphics[width=.3\textwidth]{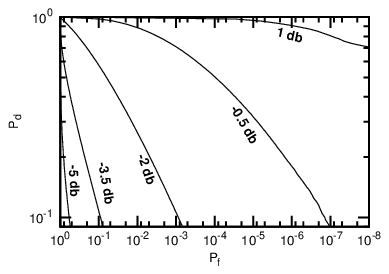}
\label{fig:analysis_alltagexp}
}
\caption{Detection curves for different choices of code and propagation model.}
\end{figure*}

Section \ref{section:TagSpotting} has described the design of tags and
introduced an universal detector\footnote{a detector for which, for
  any chosen probability of error $P_f$, the corresponding probability
  of detection $P_d$ can only increase when the SINR is increased. A
  universal detector is particularly suited to our purposes, given the
  unpredictable nature of interference power.} that does not require
an estimate of the combined power of noise and background
interference. However, the classic theoretical
results on the detection and false alarm probability distributions are
not readily applicable to our more complicated tag detector. In the
following we will analyze, using rather conservative fading models,
the performance achievable by a few particular tag families. At first we
will consider the performance achievable when searching for a single
tag signal, after which we will generalize to larger families of tags.
The analysis will make use of two fading models, a narrowband fading
model and a wideband fading model which assumes Rayleigh propagation.
Due to the limited timespan of tags, these two models 
describe short-term fading effects only.

\noindent {\bf Single Tag.} Let us consider a tag $ t $.  In what
follows, we will denote by $ c $ the number of two-carrier groups used
in the tag's construction. Assume that, for each thin carrier, the
receiver noise has power $ n $ and its distribution can be modeled by
a complex Gaussian random variable $ \mathcal{N}(0, n)$. We further
assume that for the active thin carriers the average signal power in
the frequency band corresponding to any carrier is $p$. Under the
assumption of narrowband transmission and without including the
receiver noise contribution, the received signal obtained after
demodulating one of the carriers can be modeled as a circularly
uniform complex variable of constant amplitude $ \sqrt{p} $, while in
the case of wideband transmission the signal is modeled by a complex
Gaussian random variable $\mathcal{N}(0,p)$. Assume that each thick
carrier is composed of $ \alpha $ thin carriers, out of which $ \beta
$ are active. In the case of the system presented in the previous
section $ \alpha = 8 $ and $ \beta = 4 $. Let $ P_t $ denote the set
of thin carriers activated when tag $ t $ is transmitted, $ Q_t $
represent the thin carriers that, while they are not activated, belong
to active thick carriers and $ R_t $ the thin carriers that belong to
unactivated thick carriers. In accordance to the definition given when
introducing the tag detector, we denote through $ C_t = P_t \cup Q_t $
the set of all thin carriers belonging to activated thick carriers,
regardless of whether they are active or not. Let $ r_f $ be the
complex values obtained after computing a fast Fourier transform of
the real and complex components of a sampled tag signal, i.e. the
received signal values corresponding to the various thin carriers. Let
$ || \cdot || $ denote the $ l^2 $ norm.

Under the assumption of independent Rayleigh fading, it results that
for each active thin carrier the amplitude of the received signal is
distributed according to a complex Gaussian random variable $
\mathcal{N}(0,p+n) $. Choosing a threshold value $ \gamma $ for the
quantity defined in Equation \ref{equation:tag}, we can write the
probability of detection as:

\begin{equation}
P_d = P\left(\frac{\sum_{f \in C_t} ||r_f||^2}{\sum_{f \notin C_t} ||r_f||^2}  > \frac{\gamma}{1-\gamma} 
\right)
\label{equation:pd_gamma1}
\end{equation}

or

\begin{equation}
P_d = P\left(\frac{\sum_{f \in P_t} ||r_f||^2 + \sum_{f \in Q_t} ||r_f||^2}{\sum_{f \in R_t} ||r_f||^2}  > \frac{\gamma}{1-\gamma} \right)
\label{equation:pd_gamma2}
\end{equation}

It results from the previous paragraph and due to the independence of
the circular random variables considered that the sums in the above
equations can be written using the Chi-Square distribution with $ d $
components, denoted through $ \chi^2_d $. Namely $ \frac{1}{p+n} \sum_{f
  \in P_t} ||r_f||^2 \sim \chi^2_{2 \beta c} $, $ \frac{1}{n} \sum_{f \in Q_t}
||r_f||^2 \sim \chi^2_{2 (\alpha-\beta) c} $ and $ \frac{1}{n} \sum_{f \in R_t}
||r_f||^2 \sim \chi^2_{2 \alpha c} $. 

In the case of narrowband fading, the amplitude of the received signal
for each active carrier is constant.  Therefore the first of these
sums can be written using the noncentral Chi-Square distribution
with parameter $ \lambda = 2 \beta c \frac{p}{n} $. We write $ \frac{1}{n} \sum_{f
  \in P_t} ||r_f||^2 \sim \chi^{2}_{2 \beta c}\left(2 \beta c
\frac{p}{n}\right) $.

The probability of false alarm in the case of a single tag (and a
single tested hypothesis) can be obtained by setting $ p = 0 $ in the
above formulas. Therefore, writing the ratio of the two Chi-Squared
random variables using a random variable $ f $ that follows the
Fisher-Snedcor F-distribution\cite{johnson95continuous}, $ f \sim
\mathcal{F}(2 \alpha c,2 \alpha c) $,

\begin{equation}
P_f(t) = P\left(f  > \frac{\gamma}{1-\gamma} \right)
\label{equation:pf_gamma}
\end{equation}

Figures \ref{fig:analysis_1tag} and \ref{fig:analysis_1tagexp} present
the detector's behavior at different $ SNR $ values in the case of
wideband and narrowband signals, respectively. \footnote{The SNR
  figures are computed using the power and noise figures for a thick
  carrier, that is $ SNR = \frac{\beta p}{\alpha n} $.}


\noindent {\bf Choosing the number of active carriers.}
Consider in the following a problem mentioned in Section
\ref{section:Structure}, namely the optimal number of thick carriers $
q $ that should be activated during a tag transmission in order to
maximize the performance of the detector. Let $ N $ denote the total
number of thick carriers. In order to obtain closed-form results, we
use a simplified model of tags in which we set $ \alpha = \beta $,
that is active thick carriers will use all thin subcarriers for
transmission.  Let $ f^{'} $ be a random variable generated using the
corresponding Fisher-Snedcor distribution, $ f^{'} \sim \mathcal{F}( 2
\alpha q,2 \alpha (N-q)) $.  Under this assumption we can simplify the
formulas for the probability of detection and false alarm in the
wideband case to:
$
P^{'}_d(t) = P\left(\left(1+\frac{p}{n}\right) f^{'}  >
\frac{\gamma}{1-\gamma} \right)
$
and
$
P^{'}_f(t) = P\left(f^{'}  > \frac{\gamma}{1-\gamma} \right)
$

\begin{figure}[tb]
\centering
\includegraphics[width=.4\textwidth]{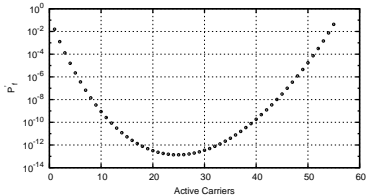}
\caption{The probability of typical false alarms as a function of the
  number of active carriers. The total number of carriers used in tag
  construction is 56.}
\label{fig:pcarriers}
\end{figure}

It results that the probability distribution of the receiver response
corresponding to detections is just a scaled version of the
probability distribution corresponding to false alarms. We introduce a
new performance measure in order to characterize the performance
change due to the choice of $ q $. Let $ \gamma_0 $ be the value of $
\gamma $ for which $ P^{'}_d = \frac{1}{2} $, at a SNR value of 0
dB. Figure \ref{fig:pcarriers} plots the behavior of $ P^{'}_f $, for
a detector with a threshold $ \gamma_0$, for different values of $ q
$, when $ N = 56 $. The quantity plotted represents the value of the
tail of the probability of false alarm in the typical detection
region.

\noindent {\bf Families of Tags.} The next point in our analysis will
be considering the situation in which the detector searches for
multiple hypotheses. Assume therefore that the tag $ t $ is a member
of a family of tags $ T $, as described in Section
\ref{section:TagSpotting}. It can be readily observed that both in the
narrowband and wideband cases the probabilities of detection remain
unchanged. The probability of false alarm can be rewritten as:

\begin{displaymath}
P_f(T) = P\left(\max_{t \in T} \left(\frac{\sum_{f \in C_t} ||r_f||^2}{\sum_{f \notin C_t} ||r_f||^2}\right)  > \frac{\gamma}{1-\gamma} \right)
\end{displaymath}

where $ ||r_f||^2 \sim \chi_2^{2} $. 

Figures \ref{fig:analysis_60tag} and \ref{fig:analysis_60tagexp}
present the detector's behavior, evaluated through Monte Carlo
simulation, at different SNRs in the case of wideband and narrowband
signals, respectively for the $ (28, 13, 60) $ code mentioned in
Section \ref{section:TagSpotting}.  In particular, Figure
\ref{fig:analysis_60tagexp} compares the theoretical predictions with
the experimental results presented in Section
\ref{section:Evaluation}. The green cross at the top of the figure
presents the experimentally measured detection probability for tags in
the presence of background noise only, at an SNR value of 1 dB, as
shown in in Figure \ref{fig:expnonoise}, while the red cross presents
the theoretical value at the same SNR. The experimental data and the
theoretical curve, which indicates a probability of detection nearing
one, are in agreement.

Figures \ref{fig:analysis_alltag} and
\ref{fig:analysis_alltagexp} illustrate the same detection curves in
the case of the simple $ (28, 1, 2^{28}) $ code.

The power levels on the carriers are being summed up in the hypotheses
in different ways. Since the variables that are summed up are chosen
from the same set, the probability that the maximum present in the
function will cross any chosen threshold $ \gamma' $ is significantly
lower than what the sum bound on the individual probabilities of
errors associated with the different codewords would predict.

The simpler quantity $\max_{t \in T} {\sum_{f \in C_t} ||r_f||^2}$ can
be bounded using an initial symmetrization step
\cite{ledoux06probability} and deriving, using generic chaining
\cite{talagrand96majorizing}, a Dudley-like inequality
\cite{dudley67thesizes} on the probability that the maximum exceeds
any given threshold. The resulting bound limits the increase of the
necessary threshold, for any fixed probability of false alarm, to a
quantity of the form $ O(log(N)) $ where $ N $ is the number of
codewords used. For reasons of space we have not included the
derivation of the bound.

For the code construction using the two-carrier groups presented in
Section \ref{section:TagSpotting}, a simple upper bound on the
probability of false alarm can be derived by considering the case of
the simplest $ (28, 1, 2^{28})$ code, which has the largest
probability of false alarm of all possible codes since it includes all
possible codewords . Consider a set of pairs of Chi-Square distributed
random variables $ (x_{i,1}, x_{i,2}) $ and let $ x^{M}_i $ and $
x^{m}_i $ represent the maximum and minimum, respectively, in each
pair. The probability of false alarm can be thus written, for the
afore mentioned code, as:

\begin{displaymath}
P_f(T) = P\left( \frac{\sum_{\forall i} x^{M}_i}{\sum_{\forall i} x^{m}_i}
 > \frac{\gamma}{1-\gamma} \right)
\end{displaymath}

The formula above has been used in order to derive the detection
curves for the code mentioned, which are presented in Figures
\ref{fig:analysis_alltag} and \ref{fig:analysis_alltagexp}.

\section{Applications}
\label{section:Applications}

\subsection{Congestion Control}
\label{section:CongestionControl}

In this section and the next one we present the use of Tag Spotting in
two applications aimed towards obtaining a fair resource allocation in
multi-hop wireless networks.
We would like to emphasize that attaining a fair distribution of
resources when using a wireless medium is, in our opinion, a
neighborhood-centric problem.
We explore the structure, the granularity and the rate of information
that hosts within a neighborhood should exchange in order to solve the
fairness problem
At the medium access layer, all hosts must be able to gain access to
the wireless channel.  At the transport layer, all flows must be able
to attain a sustained transfer rate comparable to the transfer rates
of flows with which they compete, i.e flows that share the same
wireless neighborhoods.

In a wireless setting, congestion is not always primarily experienced
by the flow that causes it and a neighborhood-wide signaling mechanism
becomes necessary.  

\begin{figure}[t]
\centering
\includegraphics[width=.34\textwidth]{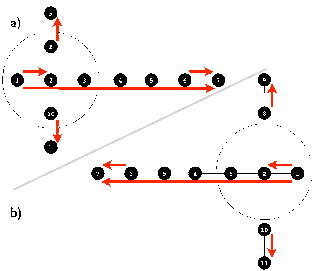}
\textcolor{white}{
\subfloat[]{
\label{fig:wcptopology}
}
\subfloat[]{
\label{fig:wcptopology2}
}
}
\caption{Chain-cross topology with all competing flows separated by at most one transmission range (upper) and with some competing flows separated by more than the transmission range (lower).}
\end{figure} 

\begin{figure*}[t]
\centering
\subfloat[for the topology in Figure \ref{fig:wcptopology}.]{
\includegraphics[width=.31\textwidth]{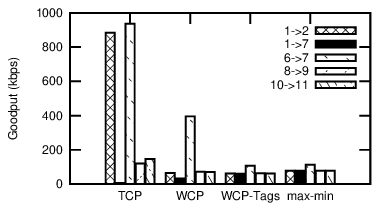}
\label{fig:wcpresults}
}
\hfill
\subfloat[for the topology in Figure \ref{fig:wcptopology2}.]{
\includegraphics[width=.31\textwidth]{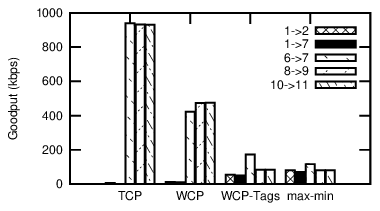}
\label{fig:wcpresults2}
}
\hfill
\subfloat[under the modified scheduling policy for the topology in Figure \ref{fig:wcptopology2}.]{
\centering
\includegraphics[width=.31\textwidth]{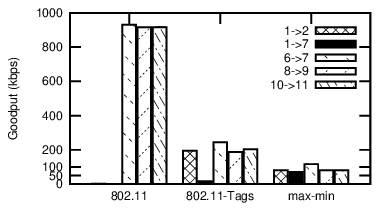}
\label{fig:schdresults}
}
\caption{Goodput Results}
\end{figure*}

With this in mind, we extend WCP~\cite{rangwala08understanding}, a
recent AIMD-based scheme from the congestion control literature, by
using Tag Spotting for communicating congestion notifications, and we
assess the achieved performance of both the original WCP protocol and
its extended derivative through simulations. In the original WCP,
there are two types of information broadcasted by each node in data
and ack packets: a congestion bit that indicates congestion events to
its neighbors and the maximum of the round trip times (RTTs) of flows
traversing it, a metric used in achieving a max-min fair allocation.
This maximum RTT is then used to pace the rate increases of the AIMD
controllers which set the rates of the flows traversing the
neighborhood.
Both loss rates and delays experienced by competing flows passing
through a congested neighborhood may vary widely, significantly more
than in wired networks. This causes the senders' AIMD
controllers to increase their rates at significantly different paces
following a congestion event, unless a common loop duration is used.
For more details on WCP, the interested reader is referred
to~\cite{rangwala08understanding}.

We have simulated the performance of WCP using the Qualnet network
simulator \cite{qualnet}.  We have extended Qualnet's physical layer
simulation in order to also handle the likelihood of tag detection
using the detection probabilities measured in Section
\ref{section:Evaluation}.  The content of tags is composed of one
congestion bit and a field that encodes, on a logarithmic scale with
base $ \sqrt[3]{2} $, the value, in milliseconds, the longest RTT
of all flows traversing the tag emitter.  We call the tag-based
implementation of WCP, WCP-Tags.  For the original WCP we have used a
broadcasting mechanism that shares the same information as WCP-Tags,
however the limit SNR for broadcast detection has been set at the same
value at which successful payload data decoding occurs, since the
original WCP broadcasts are inserted into the payload of data packets.
All hosts use the regular 802.11 MAC for ad-hoc networks with default
simulator values, with the only modification that the number of
allowed MAC layer retransmissions has been doubled from its default
value in order to decrease the rate of packet drops 
and increase the likelihood of tag reception.

Figure \ref{fig:wcptopology} illustrates a textbook configuration for
evaluating congestion control protocols in wireless environments. The
two short flows on the outside of the central chain of nodes are
within the transmission range of node 2, and we can therefore expect
that the two variants of WCP will have similar performance. Figure
\ref{fig:wcpresults} illustrates the rates obtained by TCP, original
WCP (``WCP'') and the tag-based implementation of WCP
(``WCP-Tags''). It can be readily observed that, while TCP leads to
starvation of the central flow, both WCP and WCP-Tags manage a fairer
rate allocation.

Discussing the results of these experiments requires taking into
account, above all, the fairness achieved and secondly the
throughput. 
It is well known that supporting a long flow in a wireless
multi-hop network is possible only at rates significantly lower than
the maximal link speed \cite{li01capacity}.  
Any increase in the rate of the long flow in the figure will involve a
drastic reduction in the rates of the other, shorter flows.  To make
this point more precise, we compute using brute force simulations and
the theoretical framework in \cite{ApoorvaToN} the max-min rate
allocation for these flows and compare it to the other allocations. It
is evident that both the original WCP and WCP-Tags yield rates which
are close to the max-min optimal rate allocation.

Figure \ref{fig:wcptopology2} illustrates a variation of the previous
topology in which the original WCP cannot effectively signal
congestion between the involved hosts, due to the fact that some hosts
are within the interference range but outside the data transmission
range of each other.  In particular, under the original WCP node 2
cannot inform nodes 8 and 10 that it is congested and the long flow is
almost starved, similarly to what happens under TCP.  In contrast,
WCP-Tags does not starve the long flow as nodes 8 and 10 reduce their
rates once they receive notifications through tags that node 2 is
congested.  The rate allocations achieved by the three protocols as
well as the max-min rate allocation for this topology are illustrated
in Figure \ref{fig:wcpresults2}. As before, WCP-Tags yields rates
which are close to the max-min optimal allocation.

\subsection{Scheduling}
\label{section:Scheduling}

A large class of scheduling algorithms for wireless multi-hop networks
are centered on ideas such as queue equalization using backpressure
\cite{stoylar05maximizing} \cite{chen06cross}, broadcasting local
congestion indications \cite{xu03enhancing} or creating a
computationally tractable approximation of an optimal schedule
\cite{lin04joint} \cite{jain03impact}. 
Some of the benefits of
schemes that use local communication will be illustrated in the
current section through the evaluation of a rather simple scheduling
scheme, designed as an extension to the ad-hoc mode of the the 802.11
MAC.  The simple mechanism presented here targets some of the
unfairness effects introduced by the 802.11 MAC which may lead to flow
starvation. 

Consider again the topology illustrated in Figure
\ref{fig:wcptopology2}. The results of Section
\ref{section:CongestionControl} have already shown that using a
standard 802.11 MAC in conjunction with TCP drives the longest flow in
this topology into starvation. Our solution preserves TCP as the
transport protocol but seeks to relieve such severely disadvantaged
flows by enhancing the scheduling algorithm. The key to achieving this
goal is an exchange of tags conveying a meaningful measure of
starvation, namely the average delay of the packets currently enqueued
for transmission.

The hosts observe all detected tags and decide that a host in their
neighborhood is starved for medium access whenever they receive a tag
conveying an average queueing time which is at least 32 times larger
than their own average queueing time. In this case the tag receiver
will enter a silence period of 15 milliseconds, allowing the starved
host to gain access to the channel and transmit its packets. These
numbers are not particularly optimized since our focus is to showcase
the benefits of using tags in scheduling with a very simple addition
to 802.11, rather than to provide a fully optimized and tested
solution. \vlad{However, we have tested the system with this choice of
  parameters in different scenario to ensure its stability.}  All
other medium access activity proceeds according to the normal 802.11
specification. We call this scheme 802.11-Tags. The only other
MAC-layer modification applied to both vanilla 802.11 and 802.11-Tags
consists in doubling the number of MAC-layer retries performed in case
of collision, just as in the previous section.

Interpreting the results in Figure \ref{fig:schdresults} requires
looking beyond the performance of individual links. In order for the
long flow to achieve a transmission rate on the order of tens of
kilobytes, all other flows must lower their rates far below the
hundreds of kilobytes available on individual links. As shown in the
previous section, a fair distribution of rates is associated with a
drastic decrease of overall throughput.  As expected, this simple MAC
layer modification cannot achieve a max-min fair distribution of rates
when used in conjunction with a standard AIMD rate controller like TCP
at the transport layer.

As the results in Figure \ref{fig:schdresults} show, the improved
scheduler alleviates flow starvation. The shortest of the two starved
flows in the figure reaches a rate similar to the ones of the other
three short flows. The long flow in the figure increases its achieved
rate from a level that cannot prevent connection timeouts and
interruptions to a sustained rate of about 15 kbps.

\section{Conclusion}
\label{section:Conclusion}

This paper has proposed a mechanism for sharing control information in
wireless networks, able to function in low SNR conditions and without
introducing new interference constraints. It evaluated its performance
through a theoretical analysis as well as experiments realized using
software radios. It was found that the communication scheme presented
is effective at SNR values as low as 0 dB, effectively covering the
carrier sense range of a wireless host.
 Two algorithm implementations that make use of this scheme have been
 evaluated through simulations, confirming its applicability in
 protocol design.

\bibliographystyle{elsarticle-num}
\bibliography{paper}

\end{document}